\documentclass[10pt,conference]{IEEEtran}

\usepackage{etoolbox}
\makeatletter
\def\ps@headings{%
	\def\@oddhead{\mbox{}\scriptsize\rightmark \hfil \thepage}%
	\def\@evenhead{\scriptsize\thepage \hfil \leftmark\mbox{}}%
	\def\@oddfoot{}%
	\def\@evenfoot{}}
\makeatother
\usepackage{verbatim}
\usepackage{xcolor}
\usepackage{amsfonts}
\usepackage{mathrsfs}
\usepackage{amsfonts}
\usepackage{amssymb}
\usepackage{graphicx}
\usepackage{epsfig}
\usepackage{epstopdf}
\usepackage{psfrag}
\usepackage{amsmath}
\usepackage{array}
\usepackage{cases}
\usepackage{subfigure}
\usepackage{cite,graphicx,amsmath,amssymb,color}
\usepackage{algorithmic}
\usepackage{algorithm}

\usepackage{algorithm}
\usepackage{algorithmic}
\usepackage{stmaryrd}
\usepackage{multirow}
\usepackage{subfig}
\usepackage{graphicx,times,amsmath} 

\usepackage{url}
\usepackage{enumerate}

\IEEEoverridecommandlockouts

\newtheorem{lemma}{Lemma}

\newtheorem{Exam}{Example}

\newtheorem{problem}{Problem}

\makeatletter
\renewcommand\normalsize{%
   \@setfontsize\normalsize\@xpt\@xiipt
   \abovedisplayskip 0.025\p@ \@plus0.05\p@ \@minus0.125\p@
   \abovedisplayshortskip \z@ \@plus0.075\p@
   \belowdisplayshortskip 0.15\p@ \@plus0.075\p@ \@minus0.025\p@
   \belowdisplayskip \abovedisplayskip
   \let\@listi\@listI}
\makeatother

\IEEEoverridecommandlockouts
\begin{document}
\bibliographystyle{IEEEtran}

\title{Optimal Transmission of Multi-Quality Tiled 360 VR Video by Exploiting Multicast Opportunities} 
\author{\IEEEauthorblockN{Kaixuan Long, Ying Cui, Chencheng Ye}\IEEEauthorblockA{Shanghai Jiao Tong University, China}\and\IEEEauthorblockN{Zhi Liu}\IEEEauthorblockA{Shizuoka University, Japan}}
\maketitle

\begin{abstract}
In this paper, we would like to investigate fundamental impacts of multicast opportunities on efficient transmission of a 360 VR video to multiple users in the cases with and without transcoding at each user.
We establish a novel mathematical model that reflects the impacts of multicast opportunities on the average transmission energy in both cases and the transcoding energy in the case with user transcoding, and  facilitates the optimal exploitation of transcoding-enabled multicast opportunities.
In the case without user transcoding, we optimize the transmission resource allocation to minimize the average transmission energy by exploiting natural multicast opportunities.
The problem is nonconvex. We transform it to an equivalent convex problem and obtain an optimal solution using standard convex optimization techniques.
In the case with user transcoding, we optimize the transmission resource allocation and the transmission quality level selection to minimize the weighted sum of the average transmission energy and the transcoding energy by exploiting both natural and transcoding-enabled multicast opportunities. The problem is a challenging mixed discrete-continuous optimization problem.
We transform it to a Difference of Convex (DC) programming problem and obtain a suboptimal solution using a DC algorithm.
Finally, numerical results demonstrate the importance of effective
exploitation of transcoding-enabled multicast opportunities in the case with user transcoding.
\end{abstract}

\begin{IEEEkeywords}
virtual reality, 360 video, multi-quality, convex optimization, DC programming.
\end{IEEEkeywords}

\section{Introduction}\label{section_1}

A Virtual Reality (VR) video is generated by capturing a scene of interest in every direction at the same time using omnidirectional cameras.
A captured video is stitched and warped onto a 3D sphere, and then projected to a 2D map using projection methods.
The resulting video is referred to as a 360 VR video.
The most commonly used projection method is equirectangular projection which projects a 3D sphere onto a rectangle~\cite{zink2019scalable}.
A user wearing a VR headset or Head Mounted Display (HMD) can freely watch the scene of interest in any viewing direction at any time, hence enjoying immersive viewing experience. VR has vast applications in entertainment, education, medicine, etc. It is predicted that the VR market will reach 30 billion USD by 2020~\cite{111}.

Increasing effort has been devoted to wireless transmission of 360 VR videos.
A 360 VR video is of a much larger size than a traditional video.
Thus, transmitting an entire 360 VR video brings a heavy burden to wireless networks. In addition, at any moment a user watching a 360 VR video is interested in only one viewing direction. Thus, transmitting an entire 360 VR video is also unnecessary.
To improve transmission efficiency for 360 VR videos, tiling technique is widely adopted.
Specifically, a 360 VR video is divided into smaller rectangular segments of the same size, referred to as tiles.
Suppose future field-of-views (FoVs) of a user can be successfully predicted. Then, transmitting the set of tiles covering each predicted FoV can save communications resource without degrading the user's quality of experience (QoE).
A VR user may change viewing directions from time to time. To tolerate the inaccuracy of viewing direction prediction and avoid view switch delay, the set of tiles covering the FoVs that may be watched shortly are transmitted.
In this paper, we focus on the transmission of a 360 VR video to multiple users instead of viewing direction prediction, assuming that the set of tiles to be transmitted to each user have been determined.

In our previous work~\cite{8428401,8478317}, we consider optimal transmission of a single-quality tiled 360 VR video in a TDMA system and an OFDMA system, respectively, by exploiting multicast opportunities.
In~\cite{Xie:2017:IQV:3123266.3123291,xiao2018bas,liu2018jet}, the authors consider transmission of a multi-quality tiled 360 VR video in singer-user wireless networks and focus on the optimal quality level selection for each tile to be transmitted.
The proposed solutions in~\cite{Xie:2017:IQV:3123266.3123291,xiao2018bas,liu2018jet} may not imply efficient multicast of a multi-quality tiled 360 VR video, as optimal resource sharing among users with heterogeneous channel conditions is not considered and multicast opportunities are ignored.
In~\cite{Ahmadi:2017:AMS:3126686.3126743,DBLP:journals/corr/abs-1901-02203}, the scenario of transmitting a multi-quality tiled 360 VR video to multiple users is considered, and multicast opportunities are utilized to improve transmission efficiency.
Specifically,~\cite{Ahmadi:2017:AMS:3126686.3126743} optimizes the quality level selection for each tile to be transmitted to maximize the total utility of all users under some communications resource constraints. The size of the optimization problem is unnecessarily large, as tiles are considered separately. In our previous work~\cite{DBLP:journals/corr/abs-1901-02203}, we study the optimal quality level selection to maximize the total utility of all users under communications resource constraints and quality smoothness constraints for adjacent tiles.
In contrast with~\cite{Ahmadi:2017:AMS:3126686.3126743}, in~\cite{DBLP:journals/corr/abs-1901-02203} we partition the set of tiles to be transmitted into subsets with different subsets for different user groups, and consider the optimization with respect to the subsets of tiles to effectively reduce computational complexity.
Note that~\cite{Ahmadi:2017:AMS:3126686.3126743,DBLP:journals/corr/abs-1901-02203} neglect the fact that channel conditions of users change much faster than their FoVs. Hence, the proposed single timescale solutions in~\cite{Ahmadi:2017:AMS:3126686.3126743,DBLP:journals/corr/abs-1901-02203} may not yield desired performance in practical systems.
In addition,~\cite{Ahmadi:2017:AMS:3126686.3126743,DBLP:journals/corr/abs-1901-02203} exploit only natural multicast opportunities and do not consider transcoding (converting a representation of a tile at a certain quality level to a representation at a lower quality level using transcoding tools such as FFmpeg) at the user side.
On one hand, user transcoding can create multicast opportunities, and hence save communications resource. On the other hand, transcoding at the user side consumes computation resource. How to optimally create transcoding-enabled
multicast opportunities for saving overall system resources remains an open problem.

In this paper, we would like to investigate fundamental impacts of multicast opportunities on efficient transmission of a multi-quality tiled 360 VR video to multiple users in the cases with and without transcoding at each user.
In contrast with~\cite{Ahmadi:2017:AMS:3126686.3126743,DBLP:journals/corr/abs-1901-02203}, we ensure that all tiles in a user's FoV are played at the same quality, and we consider the practical scenario where users' channel conditions and FoVs change at two timescales.
First, we introduce an elegant notation system for partitioning all tiles into subsets, each for a particular group of users, and specifying the relation between a subset of tiles and their target user group.
Then, we establish a novel mathematical model that reflects the impacts of multicast opportunities on the average transmission energy in both cases and the transcoding energy in the case with user transcoding, and facilitates the optimal exploitation of transcoding-enabled multicast opportunities.
In the case without user transcoding, we optimize the transmission resource allocation to minimize the average transmission energy by exploiting natural multicast opportunities. The problem is nonconvex. We transform it to an equivalent convex problem and obtain an optimal solution using standard convex optimization techniques.
In the case with user transcoding, we optimize the transmission resource allocation and the transmission quality level selection to minimize the weighted sum of the average transmission energy and the transcoding energy by exploiting both natural and transcoding-enabled multicast opportunities. The problem is a challenging mixed discrete-continuous optimization problem.
We transform it to a Difference of Convex (DC) programming problem and obtain a suboptimal solution using a DC algorithm.
To the best of our knowledge, this is the first work exploiting transcoding-enabled multicast opportunities for efficient transmission of a multi-quality tiled VR video to multiple users.
Finally, numerical results demonstrate the importance of effective exploitation of transcoding-enabled multicast opportunities in the case with user transcoding.

\section{System Model}\label{section_2}
\begin{figure*}[t]
\normalsize{
\centering
\subfigure{
\begin{minipage}{11cm}
\centering
\includegraphics[width=11cm]{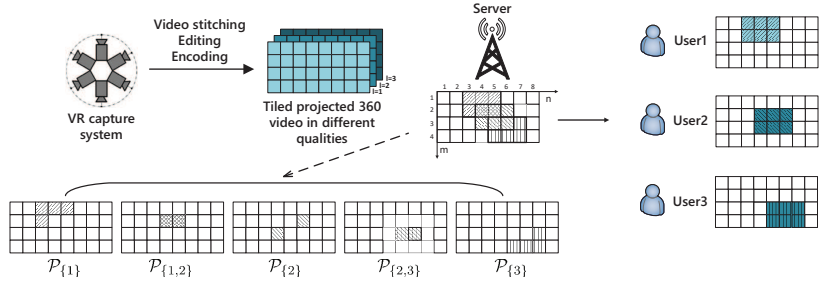}
\end{minipage}
}
\caption{  System model. $K=3$, $M=4$, $N=8$, $L=3$,
$\mathcal{G}_{1}$$=$$\{(1,3),$$(2,3),$$(1,4),$$(2,4),$$(1,5),$$(2,5)\}$, $\mathcal{G}_{2}$$=\{(2,4),$$(3,4),$$(2,5),$$(3,5),$$(2,6),$$(3,6)\}$, $\mathcal{G}_{3}$$=\{(3,5),$$(4,5),$$(3,6),$$(4,6),$$(3,7),$$(4,7)\}$,
$r_{1}=1$, $r_{2}=2$ and $r_{3}=2$. }
\label{system0}
}
\end{figure*}
As illustrated in Fig.~\ref{system0}, we consider downlink transmission of a multi-quality tiled 360 VR video from a single-antenna server (e.g., base station or access point) to $K$ $(\geq1)$ single-antenna users each wearing a VR headset in a TDMA system. Let $\mathcal{K}\triangleq\{1,...,K\}$ denote the set of user indices. A user may be interested in one viewing direction at sometime, and freely switch to another viewing direction after a while.

We consider tiling to enable flexible transmission of necessary FoVs so as to improve transmission efficiency of the 360 VR video.
Specifically, the 360 VR video is divided into $M\times N$ rectangular segments of the same size, referred to as tiles, where $M$ and $N$ represent the numbers of segments in each column and row, respectively.
Define $\mathcal{M}\triangleq\{1,...,M\}$ and $\mathcal{N}\triangleq\{1,...,N\}$. The $(m,n)$-th tile refers to the tile in the $m$-th row and the $n$-th column, for all $m\in\mathcal{M}$ and $n\in\mathcal{N}$.
Considering user heterogeneity (e.g., in cellular usage costs, display resolutions of devices, channel conditions, etc.), we pre-encode each tile into $L$ representations corresponding to $L$ quality levels using HEVC or H.264, as in Dynamic Adaptive Streaming over HTTP (DASH). Let $\mathcal{L}\triangleq\{1,...,L\}$ denote the set of quality levels. For all $l\in\mathcal{L}$, the $l$-th representation of each tile corresponds to the $l$-th lowest quality. For ease of exposition, assume that tiles with the same quality level have the same encoding rate.
The encoding rate of the $l$-th representation of a tile is denoted by $D_{l}$ (in bits/s). Note that $D_{1}<D_{2}<...<D_{L}$.

We study the system for the duration of the playback time of multiple groups of pictures (GOPs),\footnote{The duration of the playback time of one GOP is usually 0.5-1 seconds.} over which the FoV of each user does not change.
Let $r_{k}$ denote the quality requirement for user $k\in \mathcal{K}$, where
\begin{align}
r_{k}\in\mathcal{L},\quad k\in \mathcal{K}.\label{rk}
\end{align}
Note that due to the video coding structure, $\mathbf{r}\triangleq(r_{k})_{k\in\mathcal{K}}$ is fixed during the considered time duration.

To avoid view switch delay, for each user, the set of tiles that cover the FoVs that may be watched shortly will be delivered.
Let $\mathcal{G}_{k}$ denote the set of indices of the tiles that need to be transmitted to user $k$, and let $\mathcal{G}\triangleq\bigcup_{k\in\mathcal{K}}\mathcal{G}_{k}$ denote the set of indices of the tiles that need to be transmitted considering all $K$ users.
For all $\mathcal{S}\subseteq\mathcal{K}, \mathcal{S}\neq\varnothing$, let $\mathcal{P}_{\mathcal{S}}\triangleq\left(\bigcap_{k\in\mathcal{S}}\mathcal{G}_{k}\right)\bigcap\left(\mathcal{G}-\bigcup_{k\in\mathcal{K}\setminus\mathcal{S}}\mathcal{G}_{k}\right)$ denote the set of indices of the tiles that are needed by all users in $\mathcal{S}$ and are not needed by the users in $\mathcal{K}\setminus\mathcal{S}$.\footnote{Note that for all $\mathcal{S}\subseteq\mathcal{K}, \mathcal{S}\neq\varnothing$ such that $\mathcal{P}_{\mathcal{S}}\neq\varnothing$, the quality levels for the tiles in $\mathcal{P}_{\mathcal{S}}$ needed by the users in $\mathcal{S}$ may be different.}
Then $\mathcal{P}\triangleq\left\{\mathcal{P}_{\mathcal{S}}\,|\,\mathcal{P}_{\mathcal{S}}\neq\varnothing, \mathcal{S}\subseteq\mathcal{K}, \mathcal{S}\neq\varnothing\right\}$ forms a partition of $\mathcal{G}$ and $\mathcal{I}\triangleq\left\{\mathcal{S}\,|\,\mathcal{P}_{\mathcal{S}}\neq\varnothing, \mathcal{S}\subseteq\mathcal{K}, \mathcal{S}\neq\varnothing\right\}$ specifies the user sets corresponding to the partition.
In contrast with~\cite{Ahmadi:2017:AMS:3126686.3126743}, for all $\mathcal{S}\in\mathcal{I}$, we jointly consider the tiles in $\mathcal{P}_{\mathcal{S}}$ instead of treating them separately.

\begin{Exam}[Illustration of $\mathcal{P}$ and $\mathcal{I}$]\label{Exam1}
As illustrated in Fig. \ref{system0}, we consider $K=3$, $M=4$, $N=8$, $L=3$,
$\mathcal{G}_{1}$$=$$\{(1,3),$$(2,3),$$(1,4),$$(2,4),$$(1,5),$$(2,5)\}$, $\mathcal{G}_{2}$$=\{(2,4),$$(3,4),$$(2,5),$$(3,5),$$(2,6),$$(3,6)\}$, $\mathcal{G}_{3}$$=\{(3,5),$$(4,5),$$(3,6),$$(4,6),$$(3,7),$$(4,7)\}$.
Then, we have $\mathcal{P}_{\{1\}}=\{(1,3),(2,3),(1,4),(1,5)\}$, $\mathcal{P}_{\{2\}}=\{(3,4),(2,6)\}$, $\mathcal{P}_{\{3\}}=\{(4,5),(4,6),(3,7),(4,7)\}$, $\mathcal{P}_{\{1,2\}}=\{(2,4),(2,5)\}$, $\mathcal{P}_{\{2,3\}}=\{(3,5),(3,6)\}$, $\mathcal{P}_{\{1,3\}}=\varnothing$, $\mathcal{P}_{\{1,2,3\}}=\varnothing$,
$\mathcal{P}=\left\{\mathcal{P}_{\{1\}},\mathcal{P}_{\{2\}},\mathcal{P}_{\{3\}},\mathcal{P}_{\{1,2\}},\mathcal{P}_{\{2,3\}}\right\}$ and $\mathcal{I}=\left\{\{1\},\{2\},\{3\},\{1,2\},\{2,3\}\right\}$.
\end{Exam}

We consider a discrete narrowband system of bandwidth $B$ (in Hz). For an arbitrary time frame of duration $T$ (in seconds),\footnote{Note that $T$ is about 0.005 second.} let $H_{k}\in\mathcal{H}$ denote the random channel state of user $k$, representing the power of the channel between user $k$ and the server, where $\mathcal{H}$ denotes the finite channel state space. Let $\mathbf{H}\triangleq(H_{k})_{k\in\mathcal{K}}\in\mathcal{H}^{K}$ denote the random system channel state in an arbitrary time frame, where $\mathcal{H}^{K}$ represents the finite system channel state space.
We assume that the server is aware of the system channel state $\mathbf{H}$ at each time frame. Suppose the random system channel states over time frames are i.i.d.
The probability of the random system channel state $\mathbf{H}$ at each time frame being $\mathbf{h}\triangleq(h_{k})_{k\in\mathcal{K}}\in\mathcal{H}^{K}$ is given by $q_{\mathbf{H}}(\mathbf{h})\triangleq\mathrm{Pr}[\mathbf{H}=\mathbf{h}]$.

We consider TDMA.\footnote{TDMA is more analytically tractable and has applications in WiFi systems. In addition, the multicast transmission schemes and the optimization frameworks for TDMA systems can be extended to OFDMA systems, multi-user MIMO systems, etc.} Consider an arbitrary time frame.
We consider multicast. That is, one representation of each tile is transmitted at most once to serve possibly multiple users simultaneously.
The time allocated to transmit the $l$-th representations of the tiles in $\mathcal{P}_{\mathcal{S}}$ under $\mathbf{h}$, denoted by $t_{\mathbf{h},\mathcal{S},l}$, satisfies:
\begin{align}
t_{\mathbf{h},\mathcal{S},l}\geq0,\quad\mathbf{h}\in\mathcal{H}^{K},\ \mathcal{S}\in\mathcal{I},\ l\in\mathcal{L}. \label{t1}
\end{align}
In addition, we have the following total time allocation constraint under $\mathbf{h}$:
\begin{align}
\sum\nolimits_{\mathcal{S}\in\mathcal{I}}\sum\nolimits_{l\in\mathcal{L}}t_{\mathbf{h},\mathcal{S},l}\leq T,\quad \mathbf{h}\in\mathcal{H}^{K}.\label{t2}
\end{align}
The power allocated to transmit the $l$-th representations of the tiles in $\mathcal{P}_{\mathcal{S}}$ under $\mathbf{h}$, denoted by $p_{\mathbf{h},\mathcal{S},l}$, satisfies:
\begin{align}
p_{\mathbf{h},\mathcal{S},l}\geq0,\quad \mathbf{h}\in\mathcal{H}^{K},\ \mathcal{S}\in\mathcal{I},\ l\in\mathcal{L}.\label{pi}
\end{align}
The transmission energy per time frame under $\mathbf{h}$ at the server is $\sum_{\mathcal{S}\in\mathcal{I}}\sum_{l\in\mathcal{L}}t_{\mathbf{h},\mathcal{S},l}p_{\mathbf{h},\mathcal{S},l}$,
and the average transmission energy per time frame is
$\mathbb{E}\left[\sum\nolimits_{\mathcal{S}\in\mathcal{I}}\sum\nolimits_{l\in\mathcal{L}}t_{\mathbf{H},\mathcal{S},l}p_{\mathbf{H},\mathcal{S},l}\right]$,
where the expectation is taken over $\mathbf{H}\in \mathcal{H}^{K}$.
Considering joint coding over time, for all $\mathcal{S}\in\mathcal{I}$, the maximum transmission rate of the $l$-th representations of the tiles in $\mathcal{P}_{\mathcal{S}}$ to user $k\in\mathcal{S}$ is given by $\frac{B}{T}\mathbb{E}\left[t_{\mathbf{H},\mathcal{S},l}\log_{2}\left(1+\frac{p_{\mathbf{H},\mathcal{S},l}H_{k}}{n_{0}}\right)\right]$ (in bits/s), where $n_{0}$ is the power of the complex additive white Gaussian channel noise at each receiver.
To guarantee that there is no stalls during the video playback for the $l$-th representations of the tiles in $\mathcal{P}_{\mathcal{S}}$ at user $k\in\mathcal{S}$, we have the following successful transmission constraint:
\begin{align}
|\mathcal{P}_{\mathcal{S}}|D_{l}\leq \frac{B}{T}\mathbb{E}\left[t_{\mathbf{H},\mathcal{S},l}\log_{2}\left(1+\frac{p_{\mathbf{H},\mathcal{S},l}H_{k}}{n_{0}}\right)\right], \label{bxtp00}
\end{align}
where $|\mathcal{P}_{\mathcal{S}}|$ denotes the number of tiles in $\mathcal{P}_{\mathcal{S}}$.

In Section~\ref{section_3}, we consider the case where the users cannot transcode the received tiles.
In Section~\ref{section_4}, we consider the case where all users can transcode the received tiles using transcoding tools such as FFmpeg. That is, each user can convert a representation of a tile at a certain quality level to a representation at a lower quality level.
Specifically, in Section~\ref{section_3}, without considering transcoding at users, we study optimal transmission of the multi-quality tiled 360 VR video by exploiting natural multicast opportunities; and in Section~\ref{section_4}, allowing transcoding at users, we study optimal transmission of the multi-quality tiled 360 VR video by exploiting both natural and transcoding-enabled multicast opportunities.

\section{Optimal Transmission Without User Transcoding }\label{section_3}
In this section, we consider the case without user transcoding and minimize the average transmission energy under given quality requirements of all users.
In this case, for all $k\in\mathcal{K}$, the $r_{k}$-th representations of the tiles in $\mathcal{G}_{k}$ must be successfully transmitted to user $k$. For some $\mathcal{S}\in\mathcal{I}$, if there exist $k,k'\in\mathcal{S}$ such that $r_{k}=r_{k'}$, then the server can multicast the $r_{k}$-th (i.e., $r_{k'}$-th) representations of the tiles in $\mathcal{P}_{\mathcal{S}}$ to simultaneously serve user $k$ and user $k'$.
We refer to this type of multicast opportunities as natural multicast opportunities.

\begin{Exam}[Illustration of Natural Multicast Opportunities]\label{Exam2}
Consider the same setup as in Example \ref{Exam1}.
The server can unicast the first representations of the tiles in $\mathcal{P}_{\{1\}}$ and $\mathcal{P}_{\{1,2\}}$ to user 1,
the second representations of the tiles in $\mathcal{P}_{\{1,2\}}$ and $\mathcal{P}_{\{2\}}$ to user 2,
and the second representations of the tiles in $\mathcal{P}_{\{3\}}$ to user 3, separately.
In addition, in the case without user transcoding, the server can multicast the second representations of the tiles in $\mathcal{P}_{\{2,3\}}$ to user 2 and user 3, by exploiting natural multicast opportunities.
\end{Exam}

When user transcoding is not considered,
the server has to transmit the $r_{k}$-th representations of the tiles in $\mathcal{G}_{k}$ to user $k$, so that user $k$'s FoV can be played at quality level $r_{k}$. Thus, the successful transmission constraints in \eqref{bxtp00} become:
\begin{align}
|\mathcal{P}_{\mathcal{S}}|D_{r_{k}}\leq \frac{B}{T}\mathbb{E}\left[t_{\mathbf{H},\mathcal{S},r_{k}}\log_{2}\left(1+\frac{p_{\mathbf{H},\mathcal{S},r_{k}}H_{k}}{n_{0}}\right)\right],\nonumber\\
 \mathcal{S}\in\mathcal{I},\ k\in\mathcal{S}. \label{bxtp1}
\end{align}

For given quality requirements of all users $\mathbf{r}$,
we would like to optimize the transmission time $\mathbf{t}\triangleq(\mathbf{t_{h}})_{\mathbf{h}\in\mathcal{H}^{K}}$ and power $\mathbf{p}\triangleq(\mathbf{p_{h}})_{\mathbf{h}\in\mathcal{H}^{K}}$ allocation to minimize the average transmission energy 
subject to the transmission time allocation constraints in \eqref{t1}, \eqref{t2}, transmission power constraints in \eqref{pi}, and successful transmission constraints in \eqref{bxtp1}.
Specifically, for given $\mathbf{r}$, we have the following problem.

\begin{problem}[Energy Minimization without User Transcoding]\label{bP1}
\begin{align}
E^\star\triangleq\min_{\mathbf{t},\mathbf{p}} \quad&\mathbb{E}\left[\sum\nolimits_{\mathcal{S}\in\mathcal{I}}\sum\nolimits_{l\in\mathcal{L}}t_{\mathbf{H},\mathcal{S},l}p_{\mathbf{H},\mathcal{S},l}\right]\nonumber\\
\text{s.t.}\quad&
\eqref{t1}, \eqref{t2}, \eqref{pi}, \eqref{bxtp1}.\nonumber
\end{align}
Let $\left(\mathbf{t}^{\star},\mathbf{p}^{\star}\right)$ denote an optimal solution of Problem~\ref{bP1}.
\end{problem}

Problem~\ref{bP1} is nonconvex. By a change of variables, i.e., using $e_{\mathbf{h},\mathcal{S},l}\triangleq t_{\mathbf{h},\mathcal{S},l}p_{\mathbf{h},\mathcal{S},l}$ (representing the transmission energy for the $l$-th representations of the tiles in $\mathcal{P}_{\mathcal{S}}$ under $\mathbf{h}$) instead of $p_{\mathbf{h},\mathcal{S},l}$ for all $h\in\mathcal{H}^{K},\mathcal{S}\in\mathcal{I},l\in\mathcal{L}$,
we can equivalently convert Problem~\ref{bP1} to the following problem that is convex.

\begin{problem}[Convex Formulation of Problem~\ref{bP1}]\label{bP2}
\begin{align}
E^\star&\triangleq\min_{\mathbf{t},\mathbf{e}} \quad\mathbb{E}\left[\sum\nolimits_{\mathcal{S}\in\mathcal{I}}\sum\nolimits_{l\in\mathcal{L}}e_{\mathbf{H},\mathcal{S},l}\right]\nonumber\\
\text{s.t.}\quad
&\eqref{t1}, \eqref{t2},\nonumber\\
&e_{\mathbf{h},\mathcal{S},l}\geq0,\quad\mathbf{h}\in\mathcal{H}^{K},\mathcal{S}\in\mathcal{I},l\in\mathcal{L},\label{ei1}\\
&|\mathcal{P}_{\mathcal{S}}|D_{r_{k}}\leq \frac{B}{T}\mathbb{E}\left[t_{\mathbf{H},\mathcal{S},r_{k}}\log_{2}\left(1+\frac{e_{\mathbf{H},\mathcal{S},r_{k}}H_{k}}{t_{\mathbf{H},\mathcal{S},r_{k}}n_{0}}\right)\right],\nonumber\\
&\qquad\qquad\qquad\qquad\quad\qquad\mathcal{S}\in\mathcal{I},\ k\in\mathcal{S}. \label{bxte1}
\end{align}
Let $\left(\mathbf{t}^{\star},\mathbf{e}^{\star}\right)$ denote an optimal solution of Problem~\ref{bP2}.
\end{problem}

Then, we can obtain an optimal solution of Problem~\ref{bP2} using standard convex optimization techniques.

\section{Optimal Transmission With User Transcoding}\label{section_4}
In this section, we consider the case with user transcoding and minimize the weighted sum of the average transmission energy and the transcoding energy under given quality requirements of all users.
In this case, for all $k\in\mathcal{K}$, a representation of a tile in $\mathcal{G}_{k}$ of a quality level no smaller than $r_{k}$ must be successfully transmitted to user $k$. For some $\mathcal{S}\in\mathcal{I}$, if there exist $k,k'\in\mathcal{S}$ such that $r_{k}>r_{k'}$, then the server can multicast the $r_{k}$-th representations of the tiles in $\mathcal{P}_{\mathcal{S}}$ to simultaneously serve user $k$ and user $k'$, with user $k$ directly playing the received $r_{k}$-th representations of the tiles, and user $k'$ first converting the $r_{k}$-th representations to the $r_{k'}$-th representations and then playing them.
We refer to this type of multicast opportunities as transcoding-enabled multicast opportunities.
\begin{Exam}
\textit{(Illustration of Transcoding-Enabled Multicast Opportunities):}
Consider the same setup as in Example \ref{Exam1}.
The server can unicast the first representations of the tiles in $\mathcal{P}_{\{1\}}$ to user 1,
the second representations of the tiles in $\mathcal{P}_{\{2\}}$ to user 2,
and the second representations of the tiles in $\mathcal{P}_{\{3\}}$ to user 3, separately.
As in Example \ref{Exam2}, the server can multicast the second representations of the tiles in $\mathcal{P}_{\{2,3\}}$ to user 2 and user 3, by exploiting natural multicast opportunities.
Besides, the server can multicast the second representations of the tiles in $\mathcal{P}_{\{1,2\}}$ to user 1 and user 2, by exploiting transcoding-enabled multicast opportunities.
User 2 directly plays the received second representations of the tiles, and user 1 first converts the second representations to the first representations and then plays the first representations of the tiles in $\mathcal{P}_{\{1,2\}}$.
It is clear that user transcoding can create multicast opportunities, enabling more efficient transmission.
\end{Exam}

First, we establish a mathematical model to characterize the impacts of natural and transcoding-enabled multicast opportunities.
We introduce transmission quality level selection variables $\mathbf{y}\triangleq(y_{\mathcal{S},k,l})_{\mathcal{S}\in\mathcal{I},k\in\mathcal{S},l\in\mathcal{L}}$, where
\begin{align}
&y_{\mathcal{S},k,l}\in\{0,1\},\quad \mathcal{S}\in\mathcal{I},\ k\in\mathcal{S},\ l\in\mathcal{L},\label{x0}\\
&\sum\nolimits_{l\in\mathcal{L}}y_{\mathcal{S},k,l}=1,\quad \mathcal{S}\in\mathcal{I},\ k\in\mathcal{S}.\label{x1}
\end{align}
Here, $y_{\mathcal{S},k,l}=1$ indicates that the server will transmit the $l$-th representation of each tile in $\mathcal{P}_{\mathcal{S}}$ to user $k\in\mathcal{S}$, and $y_{\mathcal{S},k,l}=0$ otherwise.
\eqref{x1} ensures that the server transmits only one representation of each tile in $\mathcal{P}_{\mathcal{S}}$ to user $k\in\mathcal{S}$.
The quality level of the representation of each tile in $\mathcal{P}_{\mathcal{S}}$ to be transmitted to user $k$ is given by $\sum_{l\in\mathcal{L}}ly_{\mathcal{S},k,l}$.
With transcoding, to guarantee that user $k$' FoV can be played at quality level $r_{k}$, it is sufficient to require:
\begin{align}
\sum\nolimits_{l\in\mathcal{L}}ly_{\mathcal{S},k,l}\geq r_{k},\quad \mathcal{S}\in\mathcal{I},\ k\in \mathcal{S},\label{xiyk}
\end{align}
and the successful transmission constraints in \eqref{bxtp00} become:
\begin{align}
|\mathcal{P}_{\mathcal{S}}|D_{l}y_{\mathcal{S},k,l}\leq \frac{B}{T}\mathbb{E}\left[t_{\mathbf{H},\mathcal{S},l}\log_{2}\left(1+\frac{p_{\mathbf{H},\mathcal{S},l}H_{k}}{n_{0}}\right)\right],\nonumber\\ \mathcal{S}\in\mathcal{I},\ k\in\mathcal{S},\ l\in\mathcal{L}. \label{xtp1}
\end{align}

Besides view transmission, user transcoding also consumes energy. For ease of exposition, we assume that at each user, the transcoding energy (per time frame) for reducing the quality levels of all tiles by one are the same.
Let $E_{k}$ denote the transcoding energy (per time frame) at user $k$ for reducing the quality level of the representation of a tile by one.
Considering heterogeneous hardware conditions at different users, we allow $E_{k},k\in \mathcal{K}$ to be different.
Then, the weighted sum of the average transmission energy and the transcoding energy per time frame at all users is $\sum_{\mathcal{S}\in\mathcal{I}}\sum_{k\in\mathcal{S}}|\mathcal{P}_{\mathcal{S}}|E_{k}\left(\sum_{l\in\mathcal{L}}ly_{\mathcal{S},k,l}-r_{k}\right)$.
The weighted sum of the average transmission energy and the transcoding energy per time frame is 
$\mathbb{E}\left[\sum_{\mathcal{S}\in\mathcal{I}}\sum_{l\in\mathcal{L}}t_{\mathbf{H},\mathcal{S},l}p_{\mathbf{H},\mathcal{S},l}\right]+\beta\sum_{\mathcal{S}\in\mathcal{I}}\sum_{k\in\mathcal{S}}|\mathcal{P}_{\mathcal{S}}|E_{k}\left(\sum_{l\in\mathcal{L}}ly_{\mathcal{S},k,l}-r_{k}\right)$,
where $\beta\geq1$ is the corresponding weight factor.
Note that $\beta>1$ means imposing a higher cost on the energy consumption for user devices due to their limited battery powers.

For given quality requirements of all users $\mathbf{r}$,
we would like to optimize the transmission quality level selection  $\mathbf{y}$, transmission time allocation $\mathbf{t}$ and transmission power allocation $\mathbf{p}$ to minimize the weighted sum of the average transmission energy and the transcoding energy subject to the transmission time allocation constraints in \eqref{t1}, \eqref{t2}, transmission power constraints in \eqref{pi}, transmission quality level selection constraints in \eqref{x0}, \eqref{x1}, \eqref{xiyk}, and
successful transmission constraints in \eqref{xtp1}. Specifically, for given $\mathbf{r}$, we have the following problem.

\begin{problem}[Energy Minimization with User Transcoding]\label{P1}
\begin{align}
\overline{E}^\star\triangleq\min_{\mathbf{y},\mathbf{t},\mathbf{p}} \quad&\mathbb{E}\left[\sum\nolimits_{\mathcal{S}\in\mathcal{I}}\sum\nolimits_{l\in\mathcal{L}}t_{\mathbf{H},\mathcal{S},l}p_{\mathbf{H},\mathcal{S},l}\right]\nonumber\\
&+\beta\sum\nolimits_{\mathcal{S}\in\mathcal{I}}\sum\nolimits_{k\in\mathcal{S}}|\mathcal{P}_{\mathcal{S}}|E_{k}\left(\sum\nolimits_{l\in\mathcal{L}}ly_{\mathcal{S},k,l}-r_{k}\right)\nonumber\\
\text{s.t.}\quad&
\eqref{t1}, \eqref{t2}, \eqref{pi}, \eqref{x0}, \eqref{x1}, \eqref{xiyk}, \eqref{xtp1}.\nonumber
\end{align}
Let $(\mathbf{\overline{y}}^{\star},\mathbf{\overline{t}}^{\star},\mathbf{\overline{p}}^{\star})$ denote an optimal solution of Problem~\ref{P1}.
\end{problem}

By comparing Problem~\ref{bP1} and Problem~\ref{P1}, we can easily show the advantage of user transcoding in energy reduction, as summarized in the following lemma.
\begin{lemma}[Comparison between Problem~\ref{bP1} and Problem~\ref{P1}] \label{compare}
$\overline{E}^\star\leq{E}^\star$, where ${E}^\star$ and $\overline{E}^\star$ are the optimal values of Problem~\ref{bP1} and Problem~\ref{P1}, respectively.
\end{lemma}

Problem~\ref{P1} is a challenging mixed discrete-continuous optimization problem.
In the following, we obtain a low-complexity suboptimal solution of Problem~\ref{P1} using DC programming.

First, we convert Problem~\ref{P1} to a penalized DC problem. Specifically, by a change of variables, we use $\mathbf{e}$ instead of $\mathbf{p}$.
In addition, we equivalently convert the discrete constraints in \eqref{x0} to the following continuous constraints:
\begin{align}
&0\leq y_{\mathcal{S},k,l}\leq1, \quad \mathcal{S}\in\mathcal{I},\ k\in\mathcal{S},\ l\in\mathcal{L},\label{x2}\\
&y_{\mathcal{S},k,l}(1-y_{\mathcal{S},k,l})\leq 0, \quad \mathcal{S}\in\mathcal{I},\ k\in\mathcal{S},\ l\in\mathcal{L}.\label{x3}
\end{align}
By disregarding the constraints in \eqref{x3} and adding to the objective function a penalty for violating them, we can convert Problem~\ref{P1} to the following problem.
\begin{problem}[Penalized DC Problem of Problem~\ref{P1}]\label{P2}
\begin{align}
\min_{\mathbf{y},\mathbf{t},\mathbf{e}} \quad &\mathbb{E}\left[\sum\nolimits_{\mathcal{S}\in\mathcal{I}}\sum\nolimits_{l\in\mathcal{L}}e_{\mathbf{H},\mathcal{S},l}\right]+\rho P(\mathbf{y})\nonumber\\
&+\beta\sum\nolimits_{\mathcal{S}\in\mathcal{I}}\sum\nolimits_{k\in\mathcal{S}}|\mathcal{P}_{\mathcal{S}}|E_{k}\left(\sum\nolimits_{l\in\mathcal{L}}ly_{\mathcal{S},k,l}-r_{k}\right)\nonumber\\
\text{s.t.} \quad
&\eqref{t1},\eqref{t2},\eqref{ei1},\eqref{x1},\eqref{xiyk},\eqref{x2},\nonumber\\
&|\mathcal{P}_{\mathcal{S}}|D_{l}y_{\mathcal{S},k,l}\leq \frac{B}{T}\mathbb{E}\left[t_{\mathbf{H},\mathcal{S},l}\log_{2}\left(1+\frac{e_{\mathbf{H},\mathcal{S},l}H_{k}}{t_{\mathbf{H},\mathcal{S},l}n_{0}}\right)\right],\nonumber\\
&\qquad\qquad\qquad\qquad\qquad\mathcal{S}\in\mathcal{I},\ k\in\mathcal{S},\ l\in\mathcal{L},\label{xte1}
\end{align}
where the penalty parameter $\rho>0$ and the penalty function is given by
$P(\mathbf{y})\triangleq\sum_{\mathcal{S}\in\mathcal{I}}\sum_{k\in\mathcal{S}}\sum_{l\in\mathcal{L}}y_{\mathcal{S},k,l}(1-y_{\mathcal{S},k,l})$.
\end{problem}

Note that the objective function of Problem~\ref{P2} can be viewed as a difference of two convex functions and the feasible set of Problem~\ref{P2} is convex. Thus, Problem~\ref{P2} can be viewed as a penalized DC problem of Problem~\ref{P1}.
When the feasible set of Problem~\ref{P1} is nonempty, there exists $\rho_0>0$ such that for all $\rho>\rho_0$, Problem~\ref{P2} is equivalent to Problem~\ref{P1}, in the sense that they share the same optimal value.
We can obtain a stationary point of Problem~\ref{P2} using a DC algorithm \cite{Lipp2016}. The main idea is to iteratively solve a sequence of convex approximations of Problem~\ref{P2}, each of which is obtained by linearizing the penalty function $P(\mathbf{y})$ in the objective function of Problem~\ref{P2}. We can run the DC algorithm multiple times, each with a random initial feasible point of Problem~\ref{P2}, and select the
stationary point with the minimum weighted sum energy among those with zero penalty as the suboptimal solution of Problem~\ref{P1}.
The details are omitted due to page limitation.

\section{Numerical Results}\label{section_5}

\begin{figure}[!h]
\normalsize{
\centering
\subfigure{
\begin{minipage}{5cm}
\centering
\includegraphics[width=5cm]{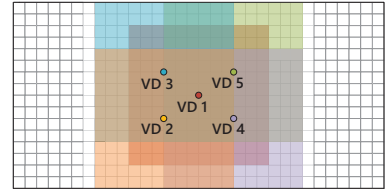}
\end{minipage}
}
\caption{ 5 possible viewing directions (VDs).}
\label{vd}
}
\end{figure}

In this section, we consider the cases with (w) and without (w/o) user transcoding and compare the proposed solutions in Section~\ref{section_3} and Section~\ref{section_4}, referred to as Proposed-w/o and Proposed-w with two baseline schemes.
In the simulation, we set $E_{k}=10^{-6}$ Joule, $B=150$ MHz,\footnote{We consider a multi-carrier TDMA with 150 channels, each with bandwidth 1 MHz.} $T=50$ms and $n_{0}=Bk_{B}T_{0}$, where $k_{B}=1.38\times10^{-23}$ Joule/Kelvin is the Boltzmann constant and $T_{0}=300$ Kelvin is the temperature.
For ease of simulation, we consider two channel states for each user, i.e., a good channel state and a bad channel state,
and set $\mathcal{H}=\{d,2d\}$, $\mathrm{Pr}[H_{k}=d]=0.5$, and $\mathrm{Pr}[H_{k}=2d]=0.5$ for all $k\in\mathcal{K}$, where $d=10^{-6}$ reflects the path loss.
For comparison, we set $\beta=1$ and evaluate the average transmission energy in the case without user transcoding and the sum of the average transmission energy and transcoding energy in the case with user transcoding.
We use Kvazaar as the 360 VR video encoder and video sequence Reframe Iran from YouTube as the video source. We set horizontal and vertical angular spans of each FoV as $100^{\circ}\times100^{\circ}$. To avoid view switch delay in the presence of view changes, besides each requested FoV we transmit an extra $10^{\circ}$ in every direction.
We set $M=18$, $N=36$ and $L=5$.
The encoding rates per tile and quantization parameters for the quality levels are shown in TABLE \ref{1}.
In addition, for ease of exposition, we consider $5$ possible viewing directions as shown in Fig. \ref{vd}, and assume that $K$ users randomly choose their viewing directions in an i.i.d. manner.
To capture the impact of the concentration of the viewing directions, assume requested viewing directions follow a Zipf distribution.
In particular, the $c$-th popular viewing direction is chosen with probability $\frac{c^{-\gamma}}{\sum_{c\in\{1,...,5\}}c^{-\gamma}}$, where $c\in\{1,...,5\}$ and $\gamma$ is the Zipf exponent.\footnote{Note that Zipf distributions are widely used to model content popularity in Internet and wireless networks. In addition, the proposed solutions are valid for arbitrary distributions of  viewing directions.} Note that a smaller $\gamma$ indicates a longer tail.
We consider 100 random choices for viewing directions of $K$ users, and evaluate the average performance over these realizations.
We assume a requested quality level follows the uniform distribution in $\{r_{lb},r_{lb}+1,...,r_{ub}\}$ with mean $\overline{r}=\frac{r_{lb}+r_{ub}}{2}$, where $r_{lb},r_{ub}\in\mathcal{L}$ and $r_{lb}<r_{ub}$.
\begin{table}[h]
\centering
\caption{\label{1} Per tile encoding rates and quantization parameters for different quality levels. }
\setlength{\tabcolsep}{2.6mm}{
\begin{tabular}{|c|c|c|c|c|c|}
\hline
Quality level &1&2&3&4&5\\
\hline
Quantization parameter &42&35&28&21&14\\
\hline
Encoding rate ($\times10^{5}$) &6.66&16.18&24.29&32.01&40.23\\
\hline
\end{tabular}}
\end{table}
\vspace{-3mm}
\begin{figure*}[t]
\normalsize{
\centering
\subfigure[ Number of users $K$ at $\gamma=0$, $r_{lb}=1$ and $r_{ub}=5$.]{
\begin{minipage}{5cm}
\centering
\includegraphics[width=5cm]{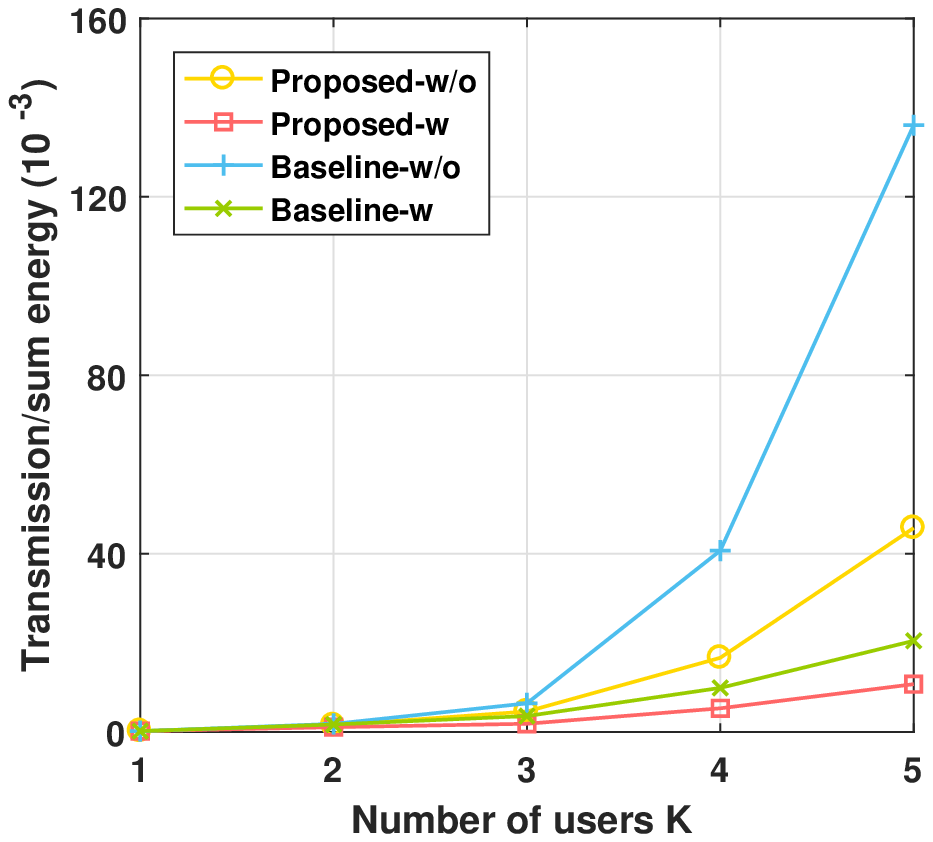}
\end{minipage}
}
\subfigure[ Zipf exponent $\gamma$ at $K=3$, $r_{lb}=1$ and $r_{ub}=5$. ]{ 
\begin{minipage}{5cm}
\centering
\includegraphics[width=5cm]{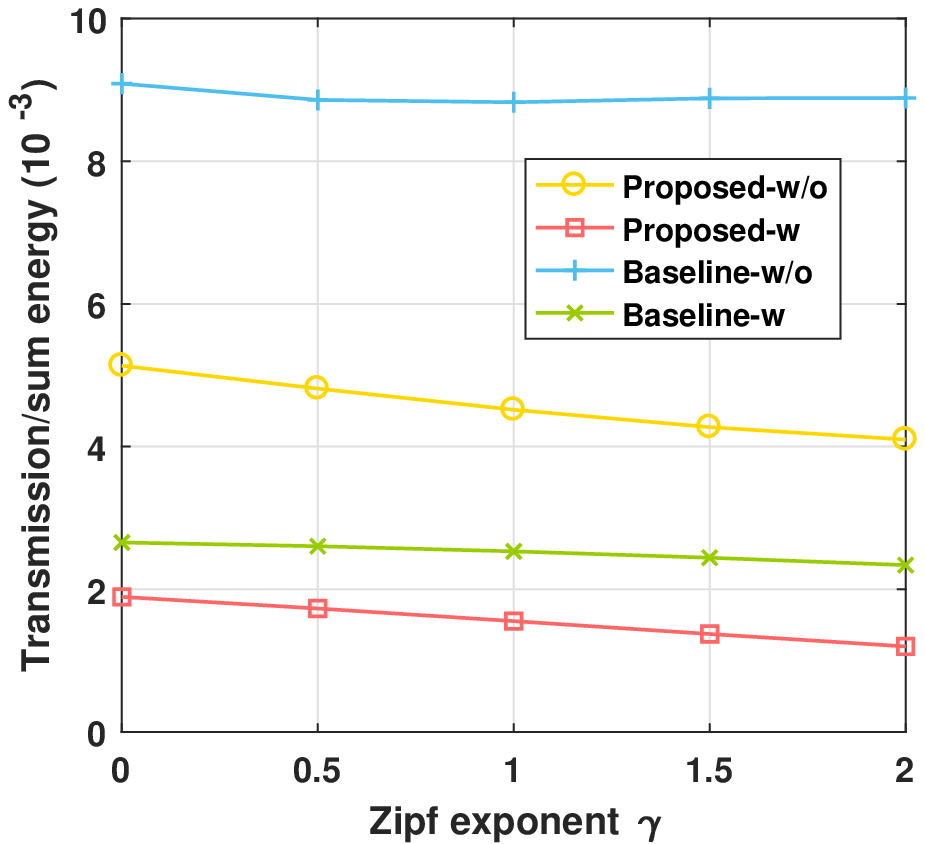}
\end{minipage}
}
\subfigure[ Mean quality level $\overline{r}$ at $\gamma=0$, $K=3$ and $r_{ub}=r_{lb}+2$. ]{ 
\begin{minipage}{5cm}
\centering
\includegraphics[width=5cm]{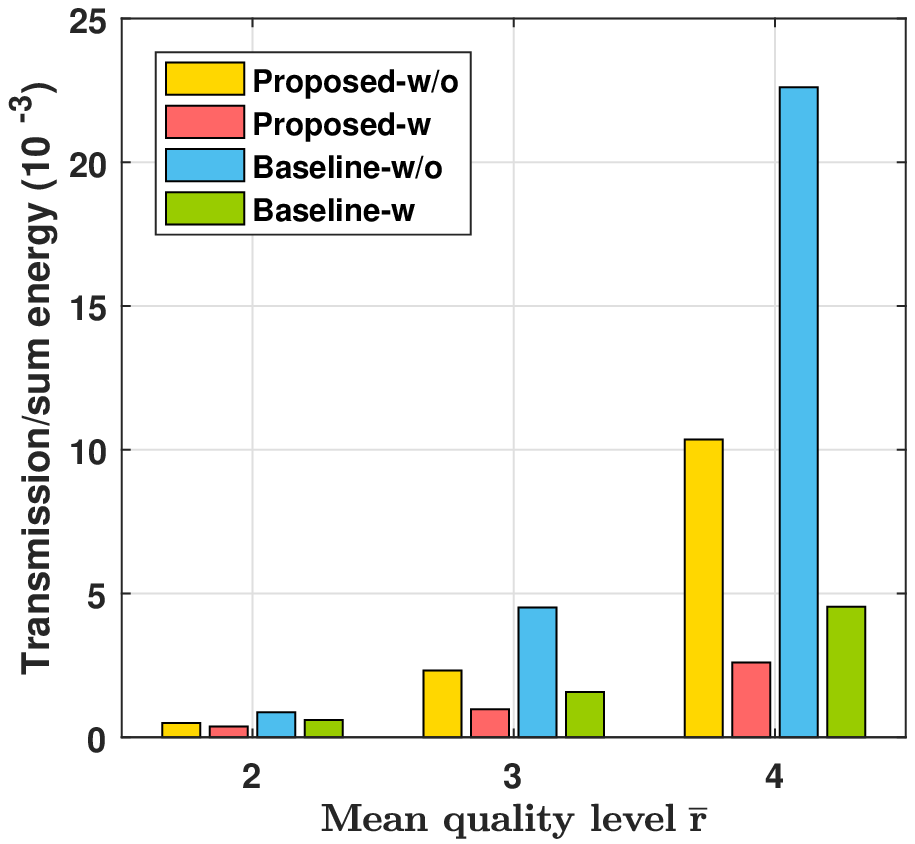}
\end{minipage}
}
\caption{Transmission/sum energy versus number of users $K$, Zipf exponent $\gamma$ and mean quality level $\overline{r}$. }
\label{sim1}
}
\end{figure*}

For comparison, we consider two baseline schemes for the cases with (w) and without (w/o) user transcoding, respectively.
In Baseline-w/o, all users are served separately using unicast no matter whether $G_{k},\:k\in\mathcal{K}$ are disjoint or not, and the corresponding optimal power and time allocation is obtained by solving a convex problem similar to Problem~\ref{bP2}.
In Baseline-w, the $r_{\mathcal{S},\max}$-th representations of the tiles in $\mathcal{P}_{\mathcal{S}}$ are transmitted to all users in $\mathcal{S}$ using multicast, where $r_{\mathcal{S},\max}\triangleq \max_{k\in\mathcal{S}}r_{k}$.
Any user $k\in\mathcal{S}$ with $r_{k}=r_{\mathcal{S},\max}$ directly plays the received $r_{\mathcal{S},\max}$-th representations of the tiles in $\mathcal{S}$, and any user $k\in\mathcal{S}$ with $r_{k}<r_{\mathcal{S},\max}$  first converts the $r_{\mathcal{S},\max}$-th representations of the tiles in $\mathcal{S}$ to the $r_{k}$-th representations, and then plays the $r_{k}$-th representations of the tiles in $\mathcal{S}$.
In Baseline-w, the corresponding optimal power and time allocation is obtained by solving Problem~\ref{P1} with $y_{\mathcal{S},k,r_{\mathcal{S},\max}}=1,\ \mathcal{S}\in\mathcal{I},\ k\in\mathcal{S}$ and $y_{\mathcal{S},k,l}=0,\ l\neq r_{\mathcal{S},\max},\ \mathcal{S}\in\mathcal{I},\ k\in\mathcal{S}$.
Note that Baseline-w/o does not utilize any multicast opportunities; Baseline-w utilize natural multicast opportunities and transcoding-enabled multicast opportunities without allowing optimal exploitation of transcoding-enabled multicast opportunities.

Fig.~\ref{sim1} illustrates the average transmission energy for the case without user transcoding and the sum of the average transmission energy and transcoding energy for the case with user transcoding versus the number of users $K$, Zipf
exponent $\gamma$ and the mean quality level $\overline{r}$. Both measurement metrics are also referred to as energy for short.
From Fig.~\ref{sim1} (a) and Fig.~\ref{sim1} (c), we can see that the energy of each scheme increases with $K$ and with $\overline{r}$, as the traffic load increases with $K$ and with $\overline{r}$. From Fig.~\ref{sim1} (b), we can see that the energies of Proposed-w/o, Proposed-w and Baseline-w decrease with $\gamma$, as these schemes exploit natural multicast opportunities that increase with $\gamma$; the energy of Baseline-w/o does not change with $\gamma$, as Baseline-w/o does not utilize natural multicast opportunities.
From Fig.~\ref{sim1}, we can see that Proposed-w/o outperforms Baseline-w/o, revealing the importance of exploiting natural multicast opportunities;
Proposed-w and Baseline-w outperform Proposed-w/o, demonstrating the importance of exploiting transcoding-enabled multicast opportunities;
Proposed-w outperforms Baseline-w, showing the importance of optimally exploiting transcoding-enabled multicast opportunities.

\section{Conclusions}\label{section_6}
In this paper, we consider optimal transmission of a 360 VR video to multiple users in the cases with and without transcoding at each user in a TDMA system.
In the case without user transcoding, we optimize the transmission resource allocation to minimize the average transmission energy by exploiting natural multicast opportunities, and obtain an optimal solution using convex optimization techniques.
In the case with user transcoding, we optimize the transmission resource allocation and the transmission quality level selection to minimize the weighted sum of the average transmission energy and the transcoding energy by exploiting both natural and transcoding-enabled multicast opportunities, and obtain a suboptimal solution using a DC algorithm.
To the best of our knowledge, this is the first work exploiting transcoding-enabled multicast opportunities for efficient transmission of a multi-quality tiled VR video to multiple users.
The proposed mechanisms and frameworks for TDMA systems can be extended to OFDMA systems, multi-user MIMO systems, etc.

\bibliography{DIP}
\end{document}